\documentstyle[12pt,epsfig]{article} 
\newlength{\dinwidth}                       
\newlength{\dinmargin}                      
\def\lsim{\mathrel{\rlap{\lower4pt\hbox{\hskip1pt$\sim$}}
    \raise1pt\hbox{$<$}}}                
\def\gsim{\mathrel{\rlap{\lower4pt\hbox{\hskip1pt$\sim$}}
    \raise1pt\hbox{$>$}}}                
\begin{document}
\sloppy
\thispagestyle{empty}
\begin{titlepage}

\begin{flushleft}
{\tt hep-ex/9609017} \\[0.1cm]
September 1996
\end{flushleft}
\vspace{2cm}
\begin{center}
\Large
{\bf  Future Measurement of the Longitudinal Proton Structure Function at HERA
} \\
\vspace{1.2cm}
  \begin{large}
L.A.T.~Bauerdick$^{a}$, A.~Glazov$^{b}$, M.~Klein$^{b}$ \\
  \end{large}
\normalsize
\vspace{1cm}
$^a$ DESY, Notkestra\ss{}e 85, D-22607 Hamburg, Germany \\
$^b$ DESY-IfH Zeuthen, Platanenallee 6, D-15738 Zeuthen, Germany \\
\vspace{8cm}
\large
{\bf Abstract}
\normalsize
\end{center}
\vspace{0.3cm}
A study is presented of a possible future measurement of the longitudinal 
structure function $F_{L}(x,Q^{2})$ with different proton beam energies 
at HERA.
\vfill 
\noindent
\small
{\it To appear in the proceedings of the workshop ``Future Physics at 
HERA'', DESY, Hamburg, 1996. }
\normalsize
 
\end{titlepage}
%
%
\mbox{}
\vspace*{1cm}
\begin{center}  \begin{Large} \begin{bf}
 Future Measurement of the Longitudinal Proton Structure Function at HERA \\
  \end{bf}  \end{Large}
  \vspace*{5mm}
  \begin{large}
L.A.T.~Bauerdick$^{a}$, A.~Glazov$^{b}$, M.~Klein$^{b}$ \\
  \end{large}
$^a$ DESY, Notkestra\ss{}e 85, D-22607 Hamburg, Germany \\
$^b$ DESY-IfH Zeuthen, Platanenallee 6, D-15738 Zeuthen, Germany
\end{center}
%
\begin{quotation}
\noindent
{\bf Abstract:}
A study is presented of a possible future measurement of the longitudinal 
structure function $F_{L}(x,Q^{2})$ with different proton beam energies 
at HERA.
\end{quotation}
%
%
\section{Introduction}
\label{sect0}
 In the one-photon exchange approximation, the deep 
inelastic inclusive scattering (DIS) cross section is given by the expression
\begin{equation}
        \frac{d\sigma}{dxdQ^2} = \frac{2\pi \alpha^2}{Q^4 x}
        \cdot [(2(1-y)+y^2) F_2(x,Q^2) -y^2 F_L(x,Q^2)].
        \label{sig}
  \end{equation}  
Here $Q^2$ is the squared four-momentum transfer, $x$ is the Bjorken 
scaling variable and $y=Q^2/sx$  the inelasticity variable with 
$s=4E_{e}E_{p}$ the centre of mass energy squared of the collision.
 The two form 
factors $F_2$ and $F_L$ are related to the cross sections $\sigma_T$ and 
$\sigma_L$ of the scattering of transversely and longitudinally polarized 
virtual photons off protons. In the Quark Parton Model $F_2$ is the sum 
of quark and antiquark distributions in the proton weighted with the 
electric quark charges squared while $F_L$ is predicted to be zero for spin 
$1/2$ partons \cite{CG}. In Quantum Chromodynamics (QCD) $F_L$ aquires a 
non zero value due to gluon radiation which is proportional to the strong 
coupling constant $\alpha_s$ \cite{AM} with possibly sizeable higher 
order corrections in QCD perturbation theory \cite{vN}. Measurements of 
$F_L$, expressed as the structure function ratio 
\begin{equation}
        R=\frac{F_L}{F_2-F_L} = \frac{\sigma_L}{\sigma_T},
        \label{R}
    \end{equation}
    were performed by various fixed target lepton-hadron scattering 
experiments at $x$ values larger than 0.01 \cite{Bo, Mil}.

A measurement of the longitudinal structure function at low $x$ at 
HERA is important for a number of reasons:
\begin{itemize}
\item{At lowest $x$
 the cross section measurement can not be uniquely 
interpreted as a determination of $F_{2}$ because the $F_{L}$ 
contribution to the cross section becomes sizeable, see eq.\ref{sig}.}
\item{A measurement of $F_{L}$ and of $F_{2}$ represents an important 
test of QCD  which uniquely describes the decomposition of the cross
section into the $F_2$ and the $F_L$ part based on a common set of
parton distributions and NLO corrections \cite{jbr}. In particular,  
the scaling violations of $F_{2}$ which at low $x$
determine the gluon distribution are predicted to be in accord with 
$F_{L}$ which is directly given by $xg$.}
\item{The lowest $Q^{2}$ behaviour of $F_{2}$ is related to $F_{L}$ or $R$: a 
hard input distribution leads to $R = 
6.2\alpha_{s}(Q^{2})/\pi$, independently of $x$ while a soft 
distribution ,which implies approximate double logarithmic scaling 
of $F_{2}$, leads to a dependence of $R$ on $\ln{1/x}$  
\cite{Y}, see also \cite{t}.}
\item{The $F_{L}$ measurements are performed at lowest possible $x$ 
where  BFKL dynamics may show up. This may not affect $F_2$ in a sizeable 
way but may lead to $F_{L}$ values 
predicted to be different by a factor of 2 from the standard DGLAP 
expectation \cite{sal}.}
\end{itemize}
 The  question  
is how precise one may hope to perform this measurement. This can be 
studied now more reliably than previously \cite{jb,mc} since the
systematics of that measurement is better defined. This paper
documents a study to measure $F_L$ with a set of 
different proton beam energies. Similar conclusions were 
reached during this workshop
 in \cite{hb}. Data at lower electron beam energies $E_e$ may be useful for systematic
 cross checks. Access to $F_L$ with lowered electron energy, however, is even more 
complicated as about two times lower scattered electron energies have to
be measured than with maximum $E_e$. Further
information on $R$ may be obtained from radiative events as originally
proposed in \cite{wk}.
\section{Cross Section Measurement}
A measurement of the longitudinal structure function requires to 
access the lowest possible scattered electron energies $E_{e}'$ 
which approximately define $y$ as $1-E_{e}'/E_{e}$. The measurement 
accuracy improves with rising $y$ like $1/y^{2}$, see eq.\ref{sig}. The kinematic 
range of the $F_{L}$ measurement is a band in the $Q^{2},x$ plane
of a $y$ interval, 
between about 0.5 and 0.85, with a low $Q^{2}$ limit given by the 
maximum accepted polar angle of the scattered electron 
$\theta_{e} \simeq 177^{o}$
and a large $Q^{2}$ limit around 100~GeV$^{2}$. A 
measurement performed at several beam energy settings appropriately 
chosen permits to cover an $x$ range at fixed $Q^{2}$ of about one 
order of magnitude. The smallest $x$ values can be covered using
highest energy data by the  
method introduced by H1 \cite{H1FL}
 which subtracts the $F_{2}$ contribution to 
the cross section assuming that $F_{2}$ is accurately described by NLO 
QCD.

The following sources of systematic errors of the high $y$ cross 
section measurement were considered:
\begin{itemize}
 \item{The uncertainty of the scattered electron energy: using the 
 kinematic peak and the $\pi_0$ mass 
reconstruction, the double angle method and Compton events 
 one may assume a scale uncertainty of 0.5\%
 which implies a cross section error of about 0.7\%
 at high $y$.}
 \item{The polar angle measurement can be as accurate as 0.5 mrad,
 even independently of the event vertex reconstruction with
 hadron tracks, 
 based on drift chambers and Silicon trackers. The resulting cross 
 section uncertainty amounts to about 0.6\%.}
 \item{The photoproduction background may cause an error of 2\%
 which assumes a 10\% control of the background. This should be 
  possible using the electron tagger systems, the hadronic backward 
  calorimeter sections  and reducing the 
 $\pi_{0}$ background part with tracking in front of calorimeters.}
 \item{At high $y$ the radiative corrections are large \cite{db} if 
 the kinematics is reconstructed with the scattered electron. These get
 reduced due to possible  track requirements or $E-p_{z}$
 cuts which allows to study  the effect of the radiative 
 corrections. Moreover, with hadron calorimetry in
 backward direction one may use as well the hadronic final state to
 reconstruct the kinematics which then is much less affected by 
 radiative effects. Altogether an uncertainty of 1\% may remain.}
 \item{Various detector and analysis efficiencies give rise to 
 an estimated uncertainty of 2\%.}
 \item{At low $E_{e}'$ the electron identification becomes difficult. 
 For $E_{e}' \geq 6.5~$GeV an error of 1\% has been achieved by 
 the H1 Collaboration \cite{H1FL}. 
 Refined cluster algorithms considering the highest energy
 cluster and the next high energy cluster can be employed and  
 information on hadron deposition in the calorimeters be used. Here 
 we assume an error of 1\%.}
 \end{itemize}
 Altogether it can be expected that a 3\% cross section error
 is achievable owing to the large statistics envisaged for this 
 measurement. This represents an improvement by a factor of 2 of the 
 H1 result obtained at $y \simeq 0.7$ with data taken in 1994.     
\section{Longitudinal Proton Structure Function $F_{L}$}
The estimated systematic cross section errors were converted into $F_L$
 measurement errors, see fig.1 (open points), which are typically 0.08
in absolute. At each $Q^2$ two or three rather precise $F_L$ measurements can be
obtained at different $x$ for the set of energies considered.
 Some of the bins are accessed with more than one beam energy
combination.
The beam energies finally chosen should include smallest and 
largest possible proton beam energies because of the measurement accuracy and $x$ range.
An important
parameter of the measurement accuracy is the minimum electron energy
$E_e'$ which was assumed to be 5~GeV. No use was made in the analysis
of a possible reduction of the $F_L$ errors by the cross calibration of the
measurement results at low $y$ where the sensitivity to $F_L$ is negligible.

At lowest $x$ information on $F_L$ can be obtained using the $F_2$
subtraction method \cite{H1FL}.
 The result of a corresponding study for the highest beam energy is
 illustrated in fig.1  (closed points). The assumptions on the cross
section error were those as described above. In the standard method
 two independent cross section measurements have to be combined.
 Here  errors have to be considered from one data set only, i.e. those
from the large and the low $y$ region. These partially are compensating 
with the exception of the electron energy miscalibration. This, however, leads to
a very distinguished departure of $F_2$ at low $y$ from any
possible QCD behaviour. Therefore it can  be
constrained further in the required QCD analysis of $F_2$ giving finally rise
to an estimated 1.5\% accuracy of the extrapolated $F_2$. Finally, 
the uncertainty of the QCD fit to 
 $F_2$ and its extrapolation to high $y$ were estimated to leave
a residual 2\% error of the subtracted $F_2$ cross section part.

 The subtraction  
method  can of course be applied to all data sets. The
data at the present HERA energies have shown already that the QCD assumption
on $F_2$ will be justified for the lower energy data since these
are limited to relatively larger $x$, at fixed $Q^2$.
The estimated $F_L$ 
errors of the subtraction method and of the data comparison 
method are similar which should enable important systematic
cross checks since the subtraction method depends on one 
energy data set only while the standard method uses at least two. These 
were not used here for any possible error reduction 
which would have been difficult to model.  
\section{Conclusions}
A measurement of the longitudinal proton structure function can be 
performed at HERA with  runs at $\geq 4$ different proton 
energies with luminosities per beam of about 10 pb$^{-1}$. Such a 
dedicated measurement series is estimated to determine $F_{L}$ for $Q^{2}$ 
values between about 4 and 100 GeV$^{2}$ with  systematic 
errors of $\simeq 0.08$ for one order of magnitude in $x$ at given 
$Q^{2}$.  This accuracy is challenging but the measurement is of fundamental 
theoretical interest.   

\newpage
\begin{figure}[b]\centering
 \begin{picture}(160,300)
\put(-150,-20){
\epsfig{file=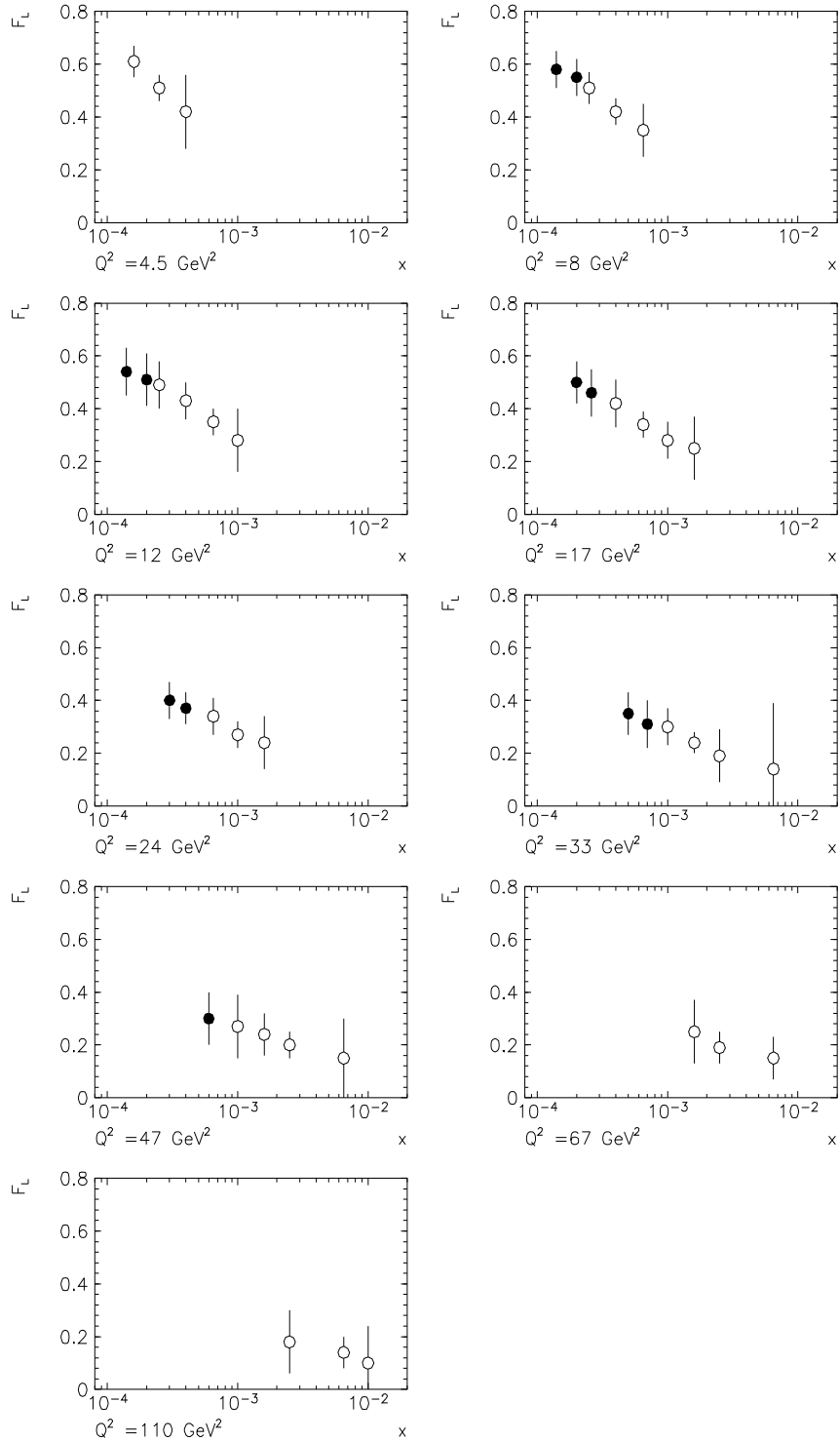,bbllx=0pt,bblly=0pt
,bburx=557pt,bbury=792pt,width=14cm,height=21cm}}
\end{picture}
  \caption{Estimated total accuracy of a measurement of the
 longitudinal structure function $F_L(x,Q^2)$ simulating data
 for an electron beam energy of $27.5$~GeV and  
 proton beam energies of 250, 350, 450 and 820 GeV
 with luminosities of
 10~pb$^{-1}$ per beam energy setting (open points). The closed points 
 at lowest $x$ represent the result of a simulation study using the
 method to determine $F_L$ after subtraction of $F_2$. These
 points are based on the highest 
 energy data set.}
        \protect\label{fls}
\end{figure}¥ 

\end{document}